\documentclass[12pt]{article}

\usepackage{ulem}

\usepackage[dvipsnames]{xcolor}
\usepackage{color}
\usepackage[english]{babel}
\usepackage{amsmath}
\usepackage{amsfonts}
\usepackage{amssymb}
\usepackage{graphicx}
\def\ai{\'{\i}}
\def\be{\begin{equation}}
\def\ee{\end{equation}}

\begin{document}
\title{\vspace{-1.5cm} \bf About the traversability of thin-shell wormholes}
\author{E. Rub\ai n de Celis\footnote{e-mail: erdec@df.uba.ar} \-  
and C. Simeone\footnote{e-mail: csimeone@df.uba.ar}\\
{\footnotesize Departamento de F\ai sica, Facultad de Ciencias Exactas y Naturales, Universidad de}\\
{\footnotesize Buenos Aires and IFIBA, CONICET, Ciudad Universitaria, Buenos Aires 1428, Argentina.}}
\date{\small \today}

\maketitle
\vspace{0.6cm} 
\begin{abstract}

Traversability in relation with tides across thin-shell wormholes is analyzed.  Conditions for a safe travel through a wormhole throat are established in terms of the parameters characterizing the geometries and reasonable assumptions regarding the travellers motion. Most convenient geometries turn to include the physically interesting example of wormholes connecting locally flat submanifolds as those associated to gauge cosmic strings. A certain relaxation of the conditions imposed and the corresponding extension of the set of admissible configurations is also briefly discussed.
\\

\vspace{1cm} 

\noindent 
PACS number(s): 4.20.-q, 04.20.Gz, 04.40.Jb

\end{abstract}

\vspace{1cm}

\section{Introduction}

Thin-shell wormholes \cite{book,visser95}  connect two submanifolds through a throat, where the matter supporting such spacetimes is located. For compact configurations the throat is a minimal area surface, while for cylindrical wormholes
it is infinite, and it admits two definitions: as a cylinder of minimal area per unit length or as a cylinder of minimal radius
 (see \cite{bronnikov} and also \cite{eisi10,cms12}). While all aspects of matter supporting thin-shell wormhole geometries as well as mechanical  stability issues have deserved a considerable amount of research, traversability in relation with tides is not a so popular subject in most literature in the area. If high curvature regions and singularities of the submanifolds joined at the throat have been removed in the cut and paste mathematical construction, the problem of tides in such spacetimes presents no novelties as long as a traveller does not reach the throat: We could always suppose that the wormhole under study connects regions of small spatial derivatives of the gravitational field, thus ensuring acceptable tides at each side (this less interesting aspect has been briefly discussed in, for example, Ref. \cite{martin}). Therefore, in principle we would not be worried about what happens between two points of an object while both of 
them are at the same side of a wormhole throat, but by the possibility of great tensions (or pressures; see below) acting on an object traveling across the throat, that is where problems associated with curvature can manifest. We will call traversable ``in practice'' a wormhole such that the relative acceleration $\Delta a$ (a precise definition is given below) between two points of an object going through the throat is physically bearable.

A reasonable condition to consider the latter would be to have a finite quotient between the tidal acceleration and the separation of the two points. At a first sight, this seems to point out that whenever an infinitely thin matter layer is present, traversability in this sense would be in trouble, as a thin-shell is associated to a discontinuity in the extrinsic curvature, which is the covariant form
of the discontinuity of the first derivatives of the metric; this seems to exclude
a finite limit for the quotient $\Delta a/\Delta x$ for two infinitely close points, one at each side of the shell. However, the problem deserves a more detailed analysis. We will begin by studying two close points of an object, one at each side of the throat, and both in the same radial direction of the geometry considered. We will then analyze the behaviour of their relative acceleration $\Delta a$ as the separation $\Delta x$ between them becomes infinitely small. 
To get a first insight of the kind of problems involved, we will first assume a rest object; 
we will afterwards consider a radial motion, as it is natural in the context of a traversability analysis.
We can understand our approach as the result of taking a shell of thickness $\delta\epsilon$ and two points at a distance 
$\Delta \eta/2$ at each side from the shell, so that 
$\Delta x = \Delta \eta+\delta\epsilon$, 
and then considering the limit in which 
$\Delta \eta \to 0$ 
so that $\Delta x\to \delta\epsilon$, with $\delta \epsilon$ the infinitely small thickness of the thin-throat. Then we will treat the case of two close points separated over a transverse direction parallel to the throat, focusing on the possible problems which appear when the object crosses the throat surface. We will find that the least problematic configurations include wormholes connecting gauge cosmic string submanifolds, as they present
metric coefficients
 $-g_{00}=g_{zz}=1$ 
 (see below) 
and therefore tides in the radial direction and in the direction parallel to the symmetry axis are not a problem; however for points along the angular direction, even for such backgrounds, the mere possibility of safe tides demands, in addition, a 
small speed. Finally, a further analysis will be briefly performed about some possible relaxation of both the conditions imposed on the geometries connected by the throat, and on the idealized infinitely-thin layer model of the matter on this surface.

\section{Preliminary analysis}

Two examples will serve to clarify the central aspects of our approach. In the first place let us analyze a thin-shell wormhole connecting two identical outer regions 
(i.e. outside the horizon radius $r_h$) of Schwarzschild geometries \cite{visser95}. At each side of the throat placed at a radial coordinate $r=r_0 > r_h$, 
the geometry has the form\footnote{We assume the convention $G=c=1$.}
\be\label{geo}
ds^2=-f(r)\, dt^2+f^{-1}(r)\, dr^2+r^2(d\theta^2+\sin^2\theta \,d\phi^2)
\ee
in the spherical Schwarzschild coordinates with $t\in \mathbb{R}$, $\theta \in (0,\pi)$, $\varphi \in [0,2\pi)$, where $r \geq r_0$ at each side, and with $f(r)=1-2M/r$ where $M$ is the mass measured at each asymptotic region. The radial acceleration of a rest particle near the throat is 
{$a^r= -M/r^2$}
(the minus sign indicating an acceleration towards the center). Given the wormhole symmetry, for a finite object slowly traversing across the throat the relative acceleration between two points, one at each side of the throat, must be twice the acceleration of one particle near the throat, that is 
\be\label{delta}
|(\Delta a)^r|=2M/r^2.
\ee
Remarkable aspects to be signaled are: 1) Because at each side the acceleration points towards the center, when extended through the wormhole throat the object is not stretched but it is compressed (a pressure force for an object). 2) The relative acceleration for one point at each side of the throat is greater by a factor 
$r/\Delta r$
when compared with the acceleration 
$(\Delta a)^r = - 2 M \Delta r/r^3$
associated with the tidal force for two nearby points at the same side of the throat and separated by 
$\Delta r$.
3) For two points, one immediately at each side of the wormhole throat, it is clear that the relative acceleration is not zero but it stays finite. According to our introductory discussion this 
analysis, though static, would imply a wormhole geometry where traversability can be jeopardized by strong tides. 

The situation can be quite different in the case of a thin-shell wormhole connecting two identical locally flat geometries, as for example those associated to gauge cosmic strings \cite{eisi04}. The spacetime geometry around such strings has the conical form \cite{vilenkin,hiscock}
\be \label{conic}
ds^2=-dt^2+dr^2+W^2r^2 \, d\varphi^2+dz^2,
\ee
with $t\in \mathbb{R}$, $\varphi \in [0,2\pi)$ and $z \in \mathbb{R}$, so that, in the associated wormhole construction, the throat is at $r=r_0$ and, at each side, we have a locally flat submanifold given by a copy of (\ref{conic}) with $r \geq r_0$, where the parameter $W \in(0,1]$ is related to the angle deficit $2\pi(1-W)$ of the conical geometry\footnote{In a gauge cosmic string spacetime given by the line element in (\ref{conic}), the angle deficit is equal to $8\pi\mu$, where $\mu$ is the mass per unit length of the string centered at the symmetry axis \cite{hiscock}.}.
The local flatness implies a vanishing acceleration for rest particles, then yielding a null relative acceleration between two points of a finite object; and this is true for both points on the same side of the wormhole throat as also for one point at each side of the throat. Hence in such spacetime no problems would appear with radial tides 
for objects crossing the throat. For points along other directions, it must be noted that the coordinates $z$ and $\varphi$ are very different, as two radial geodesics with different $z$ keep their separation constant when passing through the throat, while two radial geodesics with different $\varphi$ open up at the throat. As a consequence, no problems with tides along the $z$ direction would be expected, but tides in the angular direction are to be carefully studied (see below).

\section{Tidal acceleration - Riemann tensor}

To put these ideas in a more precise mathematical form let us recall the definition of the covariant relative acceleration \cite{book,grav}
$(\Delta a)^\mu$ which is given in terms of the Riemann tensor, the four-velocity $V^\mu$ and a vector $(\Delta x)^\mu$ which stands for the small separation of two points in spacetime:
\be
\label{A}
(\Delta a)^\mu =
- {R^\mu}_{\alpha\nu\beta} V^\alpha (\Delta x)^\nu V^\beta.
\ee   
In the following we will work with a static thin-shell wormhole space-time constructed by pasting together two copies of the same geometry at the hypersurface of the throat at $r=r_0$, these are given by the line elements 
\be\label{metric1}
ds^2 = g^{\pm}_{00}\, dt^2 + g^{\pm}_{rr}\, dr_{\pm}^2 + g^{\pm}_{\zeta\zeta}\, d\zeta^2 + g^{\pm}_{\varphi\varphi}\, d\varphi^2 \,,
\ee
where $\pm$ refers to the geometries at each side of the throat with perpendicular radial coordinates $r_{\pm} \geqslant r_0$, respectively. The non-null metric coefficients $g^{\pm}_{\mu\nu}$ depend at least on $r_{\pm}$ and, despite we are considering symmetric wormholes in which $g^{-}_{\mu\nu} = g^{+}_{\mu\nu}$, we emphasize the distinction between coefficients at each side with the $\pm$ index. The $\zeta$ coordinate represents the polar
coordinate 
$\theta \in (0,\pi)$ 
in case of spherical wormholes, or the axial coordinate 
$z \in \mathbb{R}$
in case of cylindrical geometries, 
while $t\in\mathbb{R}$ and $\varphi \in [0,2\pi)$.
The unit normal vector to the shell is defined as $n_{\mu}  = \partial_{\mu} \eta \,$, pointing from $-$ to $+$, with $\eta$ the normal coordinate associated to the radial direction such that
\be
d\eta = \pm \sqrt{g^{\pm}_{rr}} \, dr_{\pm} \,,
\ee
i.e., ($\pm$)$\eta$ measures the perpendicular proper distance at the vicinities of the throat located at $\eta = 0$, for a static observer. Considering these definitions we will generically express the diagonal metric as
\be \label{metric2}
g_{\alpha\beta}(\eta,x_{\perp})=\Theta(\eta)g^+_{\alpha\beta}(\eta,x_{\perp})+\Theta(-\eta)g^-_{\alpha\beta}(\eta,x_{\perp}),
\ee
where $x_{\perp}$ are the coordinates on the codimension-one hypersurface perpendicular to $n_\mu$ and $\Theta(.)$ is the Heaviside function. 
To begin, we study tides oven an object at rest
with a coordinate extension $\Delta r = \Theta(\eta)\Delta r_+ +\Theta(-\eta)\Delta r_-$ transverse to the throat; we then have to compute the relative acceleration between two points on the same radial direction with separation vector given by 
$(\Delta x)^\mu  = \Delta r \,{\delta_r}^\mu$. This separation will be written as
\be
(\Delta x)^\mu
= \Delta \eta \, n^\mu
\ee
with the radial proper separation $\Delta \eta$ defined as 
\be
\Delta \eta  = \sqrt{(\Delta x)^\mu (\Delta x)_\mu} = \Delta r \sqrt{g_{rr}} \,.
\ee
For a tidal acceleration in the radial direction we define, analogously, the proper tidal acceleration at the throat as $\Delta a$, such that\footnote{
{An inertial observer at rest located at the center of the symmetric throat has null proper acceleration. 
The proper tidal acceleration can be computed with the spatial components given in an orthonormal frame defined by the static observer at the 
throat. In the case of radial tides, this is the projection in the direction of $n^\mu$.}}
\be
(\Delta a)^\mu= \Delta a \, n^\mu \,.
\ee
Then, the proper magnitude of the {radial} tidal acceleration for a radially extended object is, in general,
\be \label{B}
\Delta a 
= - {R^\mu}_{\alpha \nu \beta }  \, V^\alpha \, \Delta \eta \, n^\nu \, V^ \beta\,  n_\mu \,,
\ee
and for the object at rest we have
\be
\label{.}
\Delta a 
=  {R^{r0}}_{r0} \, \Delta \eta .
\ee   
Note that the limit in which we are interested is proportional to the local behaviour of one of the components of the Riemann tensor. Any difficulty related with tides should come from the behavior of this tensor at the shell; there are no other possible problems because the wormhole construction implies pasting submanifolds without singularities, and also because the cut and paste procedure includes the condition of the continuity of the metric at the joining surface defining the wormhole throat. In order to evaluate this expression we recall the definition \cite{dau}
\be\label{Riem}
R_{r0r0}= -\frac{1}{2}(g_{rr,00}+g_{00,rr}-g_{r0,0r}-g_{0r,r0})-g_{\sigma\rho}(\Gamma^\sigma_{rr}\Gamma^\rho_{00}-\Gamma^\sigma_{r0}\Gamma^\rho_{0r}) \,,
\ee
where the derivatives respect to the radial coordinate are taken in the increasing direction of the $\eta$ coordinate, from $-$ to $+$, i.e. $\partial_r r_{\pm} =\pm1$. For metrics of diagonal form we have $g_{r0,0r}=g_{0r,r0}=0$, and as we assume static geometries we also have $g_{rr,00}=0$. 
The second radial derivative of $g_{00}$ is best expressed using (\ref{metric2}).
The definition of the $\Theta$ distribution and the continuity of the metric lead to
\be
\frac{d}{d\eta}\Theta(\pm \eta)=\pm\delta(\eta),\ \ \ \ \ \ \  g^+_{\alpha\beta}\vert_{\eta=0}=g^-_{\alpha\beta}\vert_{\eta=0}
\ee
where $\delta(\eta)$ is a Dirac delta function centered at the throat, which yields
\be\label{+-}
g_{00,r}=\Theta(\eta)g^+_{00,r}+\Theta(-\eta)g^-_{00,r}.
\ee
Note that this vanishes for the particular case $g_{00}=1$. For non-homogeneous $g_{00}$ 
at the vicinities of the shell we have
\be
g_{00,rr}
=
\left(\Theta(\eta)g^+_{00,r}+\Theta(-\eta)g^-_{00,r}\right)_{,r}
\ee
and applying once more the definition of the $\Theta$ distribution we obtain
\be
g_{00,rr}=\Theta(\eta)g^+_{00,rr}+\Theta(-\eta)g^-_{00,rr}+\delta(\eta)\left(\frac{\partial g^+_{00}}{\partial\eta}-\frac{\partial g^-_{00}}{\partial\eta}\right)n_r\,n_r,
\ee
where we have replaced $g^\pm_{00,r} = \, n_r\, \partial_\eta g^\pm_{00}$. We proceed in an analogous way to obtain the connection components:
\begin{eqnarray}
\Gamma^\sigma_{rr}&=&\Theta(\eta)\Gamma^{\sigma +}_{rr}+\Theta(-\eta)\Gamma^{\sigma -}_{rr}\nonumber\\
\Gamma^\rho_{00}&=&\Theta(\eta)\Gamma^{\rho +}_{00}+\Theta(-\eta)\Gamma^{\rho -}_{00}\nonumber\\
\Gamma^\sigma_{r0}&=&\Theta(\eta)\Gamma^{\sigma +}_{r0}+\Theta(-\eta)\Gamma^{\sigma -}_{r0}\nonumber\\
\Gamma^\rho_{0r}&=&\Theta(\eta)\Gamma^{\rho +}_{0r}+\Theta(-\eta)\Gamma^{\rho -}_{0r},
\end{eqnarray}
where for static geometries
\begin{eqnarray}
\Gamma^{\sigma\pm}_{rr}&=&\frac{1}{2}g^{\sigma\mu\pm}(g^{\pm}_{\mu r,r}+g^{\pm}_{\mu r,r}-g^{\pm}_{rr,\mu})=\frac{1}{2}g^{\sigma\mu\pm}(2g^{\pm}_{\mu r, r}-g^{\pm}_{rr,\mu}),\nonumber\\
\Gamma^{\rho\pm}_{00}&=&\frac{1}{2}g^{\rho\mu\pm}(g^{\pm}_{\mu 0,0}+g^{\pm}_{\mu 0,0}-g^{\pm}_{00,\mu})=-\frac{1}{2}g^{\rho\mu\pm}g^{\pm}_{00,\mu},\nonumber\\
\Gamma^{\sigma\pm}_{r0}&=&\frac{1}{2}g^{\sigma\mu\pm}(g^{\pm}_{\mu r,0}+g^{\pm}_{\mu 0,r}-g^{\pm}_{r0,\mu})=\frac{1}{2}g^{\sigma\mu\pm}g^\pm_{\mu 0,r},\nonumber\\
\Gamma^{\rho\pm}_{0r}&=&\frac{1}{2}g^{\rho\mu\pm}(g^{\pm}_{\mu 0,r}+g^{\pm}_{\mu r,0}-g^{\pm}_{0r,\mu})=\frac{1}{2}g^{\rho\mu\pm}g^\pm_{\mu 0,r}.
\end{eqnarray}
For the class of metrics considered here, the only non vanishing components are
\be
\Gamma^{r\pm}_{rr}=\frac{1}{2}g^{rr\pm}g^{\pm}_{rr,r},\ \ \ \ \Gamma^{r\pm}_{00}=-\frac{1}{2}g^{rr\pm}g^{\pm}_{00,r},\ \ \ \ \Gamma^{0\pm}_{r0}= \Gamma^{0\pm}_{0r}=\frac{1}{2}g^{00\pm}g^{\pm}_{00,r}.
\ee 
The components of the Riemann tensor we are interested in are written as
 \be
R_{r 0 r 0} 
=  \Theta(-\eta) R^{-}_{r 0 r 0} + \Theta(\eta) R^{+}_{r 0 r 0} 
- \delta(\eta) \kappa_{0 0} \, n_{r} n_{r} \,,
 \ee
with $R^{\mp}_{r 0 r 0}$ the smooth tensor at each side of the shell and
\be
\kappa_{0 0} = \frac{1}{2} \left(\frac{\partial g^+_{00}}{\partial \eta} - \frac{\partial g^-_{00}}{\partial\eta} \right)\Big|_{r_0}
\ee
the eigenvalue in the time-like direction of the jump in the extrinsic curvature tensor at the shell. 
To express the relative acceleration
of an object extended in the radial direction at the vicinities of the shell, and in the spirit of tidal calculations, we take $\Delta \eta$ as an infinitesimal displacement, i.e. $\Delta \eta \to d\eta$,
so that the tidal acceleration is recovered as 
\be
\Delta a = 
\int^{\Delta a/2}_{-\Delta a/2} \, da
=\int^{\Delta \eta/2}_{-\Delta \eta/2} 
- {R^\alpha}_{\beta \gamma \delta} \,  n_\alpha \, V^\beta \, n^\gamma \, V^\delta \,d\eta
\ee
by preserving a first order expansion in $\Delta \eta$ centered at the position of the shell. Specifically, to compute (\ref{.}) for an objet at rest we have
\begin{eqnarray}
\Delta a
&=&
\int^{\Delta \eta/2}_{-\Delta \eta/2} 
\,  {R^{r 0}}_{ {r}{0}} \,
\,d\eta \\
&=& 
 \left(
\int^{0}_{-\Delta \eta/2} 
\,  {{R^{r0}}_{r0}}^{-} \, d\eta + \int^{\Delta \eta/2}_{0} 
\,  {{R^{r0}}_{r0}}^{+} \, d\eta \right) 
-
\int^{\Delta \eta/2}_{-\Delta \eta/2}\, \delta(\eta) \, {\kappa^{0}}_{0}  \,d\eta
 \\
&= &
\left[  {{R^{r0}}_{r0}}^{-} +{{R^{r0}}_{r0}}^{+}\right]_{r_0} 
\frac{\Delta \eta}{2} 
+ \mathcal{O}(\Delta \eta^2) 
-
{\kappa^{0}}_{0}
\end{eqnarray}
where the expression between brackets is evaluated at the position $r=r_0$ of the shell.
The part coming from the smooth regions of the geometry is proportional to $\Delta\eta$, 
while the jump in the extrinsic curvature contributes with a fixed finite value. The relative acceleration which results is in agreement with our preliminary analysis. It reflects the discontinuous character of the gravitational field (i. e. of the first derivatives of the metric) at 
{$r_0$}, 
and according to the discussion above it would indicate that such kind of quite generic wormhole geometry, while in principle traversable, in practice would present the problem of possibly unsurmountable tides acting on a body extended across the throat due to the contribution of the extrinsic curvature jump.
In the particular case of the Schwarzschild wormhole, using that $-g^{\pm}_{00} = 1/g^{\pm}_{rr} = f(r) = 1 -2M/r$ at each side of the throat, we have
\be
{{R^{r0}}_{r0}}^{\pm}  = - \frac{f''(r)}{2} 
\ee
and
\be 
{\kappa^0}_0 
= \frac{f'(r_0)}{\sqrt{f(r_0)}} \,, 
\ee
so we obtain
\be \label{a_rad}
\Delta a
= \frac{2 M}{r_0^3} \, \Delta \eta - 
\frac{2 M}{r_0^2} \frac{1}{\sqrt{1-2M/r_0}}
\ee
or, the tidal acceleration in Schwarzschild coordinate basis $(\Delta a)^r = \Delta a \, n^r $ is
\be
(\Delta a)^r 
=
- \frac{2 M}{r_0^2} \, \left(1 - \frac{\Delta r}{r_0} \right).
\ee
To first order in $\Delta$ we have two contributions: the term proportional to $\Delta \eta$ {in (\ref{a_rad})} accounts for separations extending away from the shell into the bulk and reproduces the elongation effect produced by each smooth geometry at both sides of the shell; the other term represents a compression (negative sign) exerted by the throat, accounting for the notional extension of the object through the shell. 
The factor $g_{00}^{-1} = 1/\sqrt{f(r_0)}$ in the second term of (\ref{a_rad}) reflects the strong gravitational forces over static objects at small radii in the Schwarzschild geometry. If we consider a throat located far from the center, i.e. $r_0 \gg 2M$, we see that the condition $\Delta \eta \ll r_0$ determines that the compression exerted by the wormhole throat is always greater than the smooth tidal elongation produced by the Schwarzschild geometry.
We can not consider tides for objects smaller than the thickness of a shell with this formalism; to do so a thick-shell model would be needed. The case of vanishing tidal forces in this kind of geometry is achieved only in the limiting case with $M=0$, for which $g_{00}$ is homogeneous.

We can now consider the physically more interesting case of an object moving radially across the throat with four-velocity $V^\mu = (V^0, V^r, 0, 0)$ and separation 
$(\Delta x)^\mu = \Delta \tilde{\eta} N^\mu$ with $\Delta \tilde{\eta}$ as the proper radial separation between points of the moving object. The space-like vector $N^\mu = (N^0, N^r, 0, 0)$ is such that $N^\mu\, N_\mu = 1$ and $V^\mu\, N_\mu = 0$. 
From (\ref{A}) we note that for a radially extended object there are no tides in the perpendicular directions. Replacing as in (\ref{B}), we have the same result of (\ref{.}), 
\be \label{Bmov}
\Delta a =
(\Delta a)^\mu \, N_\mu
=
{R^{r0}}_{r0} \, \Delta \tilde{\eta} \, ,
\ee
and so the same proper radial tide of (\ref{a_rad}) is obtained for two radially separated points of the object moving radially through the throat of the wormhole constructed with the Schwarzschild geometry.
If, for example, we put $M = 10 M_{\oplus}$ and $r_0 = R_{\oplus}$, where $M_{\oplus}$ and  $R_{\oplus}$ are the mass and radius of the earth, a $1$ meter object would experience a safe travel near the vicinities of the throat with $\Delta a%^{smooth} 
\sim 10^{-6} \, \mbox{g}$, but undergo $\Delta a%^{throat} 
\sim 20\, \mbox{g}$ while traveling across the throat, where $\mbox{g} = 9.8$m/s$^2$ (particularly, this tidal acceleration would compromise the safe travel of a human being).

As it was suggested by the preliminary analysis, the situation is different at the throat of wormholes where flat or locally flat submanifolds are joined by a thin layer. Effectively, in such a case (or more generally in any case in which $g_{00}$ {is uniform} at the bulk of each side) we start from an everywhere null radial derivative of $g_{00}$, and no divergence appears in the corresponding component of the Riemann tensor. At most, we could have a finite contribution coming from the radial derivative of $g_{rr}$ if the geometry is not (locally) flat. Hence in the limit $\Delta \eta \to 0$ the relative acceleration $\Delta a$ vanishes, which reflects that the acceleration of a point particle is a continuous function across the throat, as it is null everywhere. Then such kind of wormhole would not present tidal problems in the radial direction for objects going through its throat.  

The same procedure can be applied in the case of two points separated along a direction parallel to the throat. If, to begin the analysis, we place a rest object in an equatorial plane ($\theta_0 = \pi/2$) of a spherically symmetric wormhole, 
or $z=z_0$ of a cylindrical wormhole, and extended along the azimuthal angle, we have $(\Delta x)^\mu = \Delta x_{\perp} \,e_{\perp}^{\mu}$, with $e_{\perp}^{\mu} = \delta^\mu_\varphi / \sqrt{g_{\varphi \varphi}}$ and $\Delta x_{\perp} = r_0 \, \Delta \varphi %\, \sin\theta_0 
$. The proper tidal acceleration is in the same angular direction of the extended object and defined as $\Delta a_{\perp} = (\Delta a)_\mu \,e_{\perp}^{\mu}$. If it is at rest at the shell's hypersurface we obtain
\be\label{trans0}
\Delta a_{\perp}
= {R^{\varphi 0}}_{\varphi 0} \, \Delta x_{\perp}
.
\ee
If we reproduce the calculations following Eq. (\ref{Riem}), we obtain no divergencies and the simple expression
\be
R_{\varphi 0 \varphi 0}
=
-\frac{g_{\varphi\varphi,r} \, g_{00,r}}{4\, g_{rr} } \, 
\ee
where each metric derivative can be decomposed as in Eq. (\ref{+-}). If the submanifolds connected at the throat have {uniform} $g_{00}$, as in the case of the locally flat gauge cosmic string metrics, we immediately have $R_{{\varphi} {0} {\varphi} {0}}=0$ and there are no {angular} tides at the throat; in the spherically symmetric Schwarzschild wormhole we would obtain ${R^{\varphi 0}}_{\varphi 0}=-M/r_0^3$ and the tidal force could be controlled by suitably choosing the mass and the wormhole throat radius.

If, in the more general situation, an object is moving radially across the throat with four-velocity $ dx^\mu/d\tau %V^\mu  
= (V^0, V^r, 0, 0) $, the proper tidal acceleration is given by
\be
\label{perp_mov}
\Delta a_{\perp} 
=
- \left( 
{R^{\varphi}}_{0 \varphi 0} \, V^0\, V^0 
+ 
{R^{\varphi}}_{r \varphi r} \, V^r\, V^r \, 
\right)
\Delta x_{\perp}
\ee
evaluated at the trajectory  $x^\mu(\tau)$ 
of the object. 
As before, there are no problems with divergencies coming from the $R_{\varphi 0\varphi 0}$ term. However, while in 
\be\label{Riemphi}
R_{\varphi r\varphi r}= -\frac{1}{2}(g_{\varphi\varphi,rr}+g_{rr,\varphi\varphi}-g_{\varphi r,r \varphi}-g_{r\varphi,\varphi r})-g_{\sigma\rho}(\Gamma^\sigma_{\varphi\varphi}\Gamma^\rho_{rr}-\Gamma^\sigma_{\varphi r}\Gamma^\rho_{r\varphi}) 
\ee
there are no difficulties associated to the Christoffel symbols, the second radial derivative does provide a divergent contribution, as 
\be
g_{\varphi\varphi,rr}=\Theta(\eta)g^+_{\varphi\varphi,rr}+\Theta(-\eta)g^-_{\varphi\varphi,rr}+
2\, \delta(\eta) \, \kappa_{\varphi \varphi} \, n_r\,n_r
\ee
with
\be
\kappa_{\varphi \varphi}= \frac{1}{2} \left(\frac{\partial g^+_{\varphi\varphi}}{\partial\eta}-\frac{\partial g^-_{\varphi\varphi}}{\partial\eta}\right)\Big|_{r_0}
\ee
the eigenvalue in the angular direction of the jump in the extrinsic curvature tensor at the shell.
Now, differing from the case of the radial tide where the central role was played by the $g_{00}$ component, the situation is dictated by $g_{\varphi\varphi}$. The central difference is that the radial first derivative of the $g_{\varphi\varphi}$ component can not vanish, a second derivative including a Dirac delta distribution appears, and a divergent angular tide cannot be avoided. This divergence is present in general, even in the particular case of a locally flat background.
%} 
Replacing in (\ref{perp_mov}) for a static and symmetric wormhole with line element as in (\ref{metric1}), the tidal acceleration is given as
\be \label{D_a_perp_gen}
\Delta a_\perp  %\equiv \Delta a_\mu e_\perp^\mu 
= \Delta a_\perp^{finite} + \Delta a_\perp^{div}
\ee
where 
\be\label{smooth}
\Delta a_\perp^{finite}
\equiv
\Delta a_\perp^{+}
+
\Delta a_\perp^{-}
\ee
with
\begin{eqnarray} 
\Delta a_\perp^{\pm}
&=&
\frac{\Delta x_{\perp}}{2}\,
\left[
{R^{\varphi 0}}_{\varphi0}
+
V^r \, V^r \,
 g_{rr}\, \left(
 {R^{\varphi0 }}_{\varphi0} -  \, {R^{\varphi r}}_{ \varphi r}
\right)
\right]^{\pm}_{x^\mu(\tau)}
\\
&=&
-
\frac{\Delta x_{\perp}}{8 g_{\varphi \varphi}} \,
\left[
\frac{g_{00,r}\, g_{\varphi\varphi,r} }{g_{00}\, g_{rr}}
+
V^r \, V^r \,
\left(
 g_{\varphi\varphi,r} \,
\frac{ \left( g_{00}\,g_{rr}\, g_{\varphi\varphi} \right)_{,r} }{g_{00}\,g_{rr}\, g_{\varphi\varphi}}
- 2 g_{\varphi \varphi, rr}
\right)
\right]^{\pm}_{x^\mu(\tau)}
\end{eqnarray}
the smooth part, given by the metric coefficients $g_{\mu \nu}^{\pm}$ from the geometries at each side of the throat and evaluated at the position ${x^\mu(\tau)}$ of the object, and
\be \label{div_gen}
\Delta a_\perp^{div}
\equiv
\Delta x_{\perp}\,
\delta({\eta}) 
\,{\kappa^{\varphi}}_{\varphi}
\,  g_{rr} \, 
V^r \, V^r   \big|_{x^\mu(\tau)}
\ee 
the divergent part. 
To interprete the result at the throat $r=r_0$ avoiding the delta function which is centered at the position of the infinitely-thin shell, we compute the proper tidal acceleration averaged over some notional proper time interval $\delta{\tau}$ elapsed across the trajectory through the throat, i.e.
\be
\Delta a_\perp = 
\frac{1}{\delta\tau}\, \int^{\tau_0 +\delta\tau/2}_{\tau_0 -\delta\tau/2} \Delta a_\perp \, d\tau \,.
\ee
The first term in (\ref{D_a_perp_gen}), corresponding to the finite contribution from the smooth part of the geometry at each side, remains unaltered working to first order in $\delta\tau$: this is (\ref{smooth}) evaluated at $r=r_0$. The divergent term, (\ref{div_gen}),
can be integrated by writing $V^r = dr/d\tau = g_{rr}^{-1/2}d\eta/d\tau$, to obtain
\begin{eqnarray} \label{Daperp_rad}
(\Delta a_\perp)^{div}  
&=&
\frac{{\kappa^{\varphi}}_{\varphi}}{\delta\tau}\, 
\sqrt{g_{rr}} \,V^r \,
\big|_{r_0} \,\Delta x_\perp \,.
\end{eqnarray}
We got rid of the delta function but now this term appears regulated by the inverse of the infinitely-short traveling proper time $\delta\tau$ of the object across the shell. 
Finally we can use the velocity $V^{\hat{\mu}} = dx^{\hat{\mu}}/d\tau=(\gamma, \gamma \beta, 0, 0)$ as measured in an orthonormal frame $\{ \vec{e}_{\hat{\mu}}\}$ at rest at the vicinities of the throat to
express the result using the parameters $\gamma = 1/\sqrt{1-\beta^2}$ and radial speed $\beta$ of the object as measured in the orthonormal frame. The coordinate radial component of the velocity is generically given by $V^r = V^{\hat{r}} \, (\vec{e}_{\hat{r}})^r = \gamma \beta/\sqrt{g_{rr}}$, so at the throat of the symmetric wormhole we have
\be 
\Delta a_\perp^{finite}
=
\Delta x_{\perp} \,
\left[
{R^{\varphi 0}}_{\varphi0}
+
\gamma^2 \beta^2 \,
 \left(
 {R^{\varphi0 }}_{\varphi0} -  \, {R^{\varphi r}}_{ \varphi r}
\right)
\right]^{\pm}_{r_0}
\ee
and
\be
(\Delta a_\perp)^{div}  
=
\Delta x_\perp\, \frac{\gamma \beta \,
{\kappa^{\varphi}}_{\varphi}}{\delta\tau}\,.
\ee
Both contributions are proportional to the transverse extension $\Delta x_{\perp}$ of the object. The finite term is a compression effect produced by the smooth part of the geometry, while the divergent term represents a stretching exerted by the jump in the extrinsic curvature components at the throat. Only for negligible velocities, or null curvature jump at the throat, the divergent term would be avoided. 

For the case of the thin-shell Schwarzschild wormhole, the explicit calculations yields 
\be
{{R^{\varphi 0}}_{ 0 \varphi 0 }}^{\pm}
=
{{R^{\varphi r}}_{ r \varphi r }}^{\pm}
= 
-\frac{f'(r)}{2 r  } \,,
\ee
\be
{\kappa^{\varphi}}_{\varphi} = \frac{2}{r_0} \sqrt{f(r_0)} \,,
\ee
and then 
the proper tidal acceleration for the two points on a direction parallel to the throat and traversing radially is
\be\label{dtau}
\Delta a_{\perp}
= 
\Delta x_{\perp} 
\left(
 \frac{2 \,\gamma \beta \sqrt{1-2M/r_0} }{ r_0\, \delta\tau}
- \frac{M}{r_0^3}
\right)\,.
\ee
Being $\delta\tau$ infinitely-short, we can express the result in terms of some infinitely-small size $\delta\epsilon$ associated to the radial thickness of the shell   by identifying $\gamma \beta %= (d\hat{t}/d\tau) \, (d\hat{r}/d\hat{t}) 
= dx^{\hat{r}}/d\tau
%= d\hat{r}/d\tau 
\to \delta\epsilon/\delta\tau$, i.e.
\be\label{deta}
\Delta a_{\perp} 
= 
\Delta x_{\perp} 
\left(
  \frac{2 \,\gamma^2 \beta^2 \sqrt{1-2M/r_0} }{ r_0\, \delta\epsilon}
- \frac{M}{r_0^3}  %\, \gamma^2 \left( 1 -  \beta^2 \right) \,  
\right)\,.
\ee 
We see that in the limits $\delta\tau\to 0$, or $\delta\epsilon\to 0$, corresponding to a negligible thickness of the throat, a huge transverse tide is unavoidable for the case of a moving object with some finite extent $\Delta x_{\perp}$. Nevertheless, if the object is allowed to react, the duration of the tidal acceleration would produce a finite relative velocity per unit length $\Delta v_{\perp} / \Delta x_{\perp} = \gamma \beta \, {\kappa^\varphi}_\varphi$ due to the passage through the throat or, alternatively, a finite increase in the internal energy of the body. In any case, this can be kept tolerable as long as the speed is small enough.

In the particular case of axisymmetric thin-shell wormholes, like those associated to cosmic strings, the two directions parallel to the wormhole throat are no more equivalent; then the behavior of tides in each direction can in general be quite different. The tide along the direction parallel to the symmetry axis for a radially moving object is determined by the Riemann tensor components $R_{z0z0}$ and $R_{zrzr}$. Within this framework,  wormholes connecting gauge cosmic string submanifolds turn to be, once again, of particular interest because besides a uniform $g_{00}$ component they also have $g_{zz}=1$. Then the same steps followed above for the angular direction but substituting $\varphi$ by $z$ straightforwardly show that no divergences appear, and for such backgrounds there are no problems associated with tides along the symmetry axis.

\section{Discussion}

The present discussion  seems to indicate that, under the practical approach adopted here, traversability problems given by strong {radial} tides at the throat of thin-shell wormholes could be safely avoided {for rest or radially moving objects} only in those configurations connecting submanifolds with uniform $g_{00}$. Besides, in the case of axisymmetric wormholes, tides along the longitudinal direction are not a problem for geometries with uniform $g_{zz}$. This includes the physically interesting case of gauge cosmic string thin-shell wormholes, which satisfy both conditions. For rest objects, also angular tides vanish in such configurations, and could be controlled in other backgrounds; instead, there is no way to avoid angular tides on moving objects across the thin-shell, for any background. 
In short, for finite objects in radial motion across the thin-shell, traversability in practice demands: null or small extrinsic curvature jump in the case of radially extended objects, in order to have a tolerable tidal acceleration; and null or controlable product of the speed and the transverse extrinsic curvature jump in the case of objects extended parallel to the throat, in order to produce a small enough relative velocity per unit length during the passage through the throat.

More general configurations associated to a certain relaxation of the conditions imposed on the geometries connected and also on the character of the matter layers at the throat could be considered to be admissible, but a fine tuning of the parameters
or the shape of the thin-shell throat\footnote{This refers to a class of non-symmetric solutions as, for example, wormholes with a cubic thin-shell throat composed of flat planes (the faces) where the extrinsic curvature jump vanishes, see reference \cite{book}.}, 
would be required. We can adopt the less restrictive requirement of a sufficiently small acceleration $a$ at each side of a shell of finite though little thickness $\epsilon$, so that the quotient $\Delta a/\epsilon$ is bearable (in this picture the physics within the matter layer are not considered in detail, so then only what happens immediately outside is relevant in the analysis). This approximation would be in the line of the ``traversability in practice'' condition adopted in, for instance, Ref. \cite{book} for wormholes which are not of the thin-shell class, where a maximum quotient 
$\Delta a/\ell$ is admitted if an object can withstand a maximum tidal acceleration $\Delta a$ between two points separated by a distance $\ell$.  
The condition of a little but not vanishing acceleration at each side could be achieved, for example, in the case of the spherically symmetric thin-shell wormhole connecting two Schwarzschild  geometries. One should play with the mass $M$, the throat radius $r_0$ and the thickness $\epsilon$ in order to render the quotient $\Delta a/\epsilon \sim 2M/({\epsilon} \, r^2_0)$ acceptable. We can agree that to assume a shell to be ``thin'' we should at least demand that ${\epsilon}\ll r_0$; thus the condition $\Delta a/{\epsilon}= 2M/r_0^3$ for acceptable radial tides in a motion restricted to one side of the wormhole is not enough to ensure a safe passage to the other side, because within our assumptions that quotient is much smaller than $2M/({\epsilon} \, r^2_0)$. Then we should add the requirement of a small enough mass $M$. This is in agreement with the limit $M\to 0$ found in the infinitely thin layer framework of the preceding section.

This kind of approach could also be adapted to address the problem of the transverse tides at the throat in the case of a speed that can not be neglected, which appears to be the most difficult situation (except in the particular case of the longitudinal direction in gauge cosmic string wormholes and other possible backgrounds with constant $g_{zz}$). If a finite thickness $\epsilon$ is assumed, then a specific relation between the speed $\beta$ and $\epsilon$ (see Eq. (\ref{deta})), or between $\beta$ and a finite proper time of traveling across the shell (see Eq. (\ref{dtau})), can be established in order to render admissible the relative acceleration between two points along a transverse direction.

As a final note, we should stress that the traversability analysis of the present work can also be carried out for topologically trivial configurations with shells separating inner from outer regions (see for instance \cite{gravs,em16,nos18,nos19}), as well as for multi-layer distributions in cylindrical spacetimes (see for example Refs. \cite{k13,k15,nos20}).

\section*{Acknowledgments}

We wish to thank M. C. Tomasini for questions triggering the present analysis, and E. F. Eiroa for interesting discussions.

\end{document}